\documentclass[review]{elsarticle}

\usepackage[T1]{fontenc}
\usepackage{mathrsfs}
\usepackage{xcolor}

\usepackage{lineno,hyperref}
\modulolinenumbers[5]

\journal{Spectrochimica Acta A}

\begin{document}

\begin{frontmatter}

\title{On the 3$^{1}\Pi_{u}$ state in caesium dimer}

\author[PAN]{Jacek Szczepkowski}
\author[PAN]{Anna Grochola}
\author[PAN]{Wlodzimierz Jastrzebski \corref{mycorrespondingauthor1}}
\ead{jastr@ifpan.edu.pl}
\author[FUW]{Pawel Kowalczyk
\corref{mycorrespondingauthor}}
\ead{Pawel.Kowalczyk@fuw.edu.pl}
\cortext[mycorrespondingauthor1]{Corresponding author}
\cortext[mycorrespondingauthor]{Corresponding author}
\address[PAN]{Institute of Physics, Polish Academy of Sciences,
al.~Lotnik\'{o}w~32/46, 02-668~Warsaw, Poland}
\address[FUW]{Institute of Experimental Physics, Faculty of Physics,
University of Warsaw, ul.~Pasteura~5, 02-093~Warszawa, Poland}

\begin{abstract}
Polarisation labelling spectroscopy technique was employed to study the 3$^{1}\Pi_{u}$ state of Cs$_2$ molecule. The main equlibrium constants are
$T_e=20684.60$~cm$^{-1}$, $\omega_e=30.61$~cm$^{-1}$ and $R_e=5.27$~\AA. Vibrational levels $v=4-35$ of the 3$^{1}\Pi_{u}$ state were found to be subject to strong perturbations by the neighbouring electronic states. Energies of 3094 rovibronic levels of the perturbed complex were determined.

\end{abstract}

\begin{keyword}
laser spectroscopy \sep alkali dimers \sep electronic states \sep
perturbations 
\PACS 31.50.Df \sep  33.20.Kf \sep 33.20.Vq \sep 42.62.Fi
\end{keyword}

\end{frontmatter}

\date{\today}


\newpage

\begin{figure}
	\includegraphics[width=1.3\linewidth]{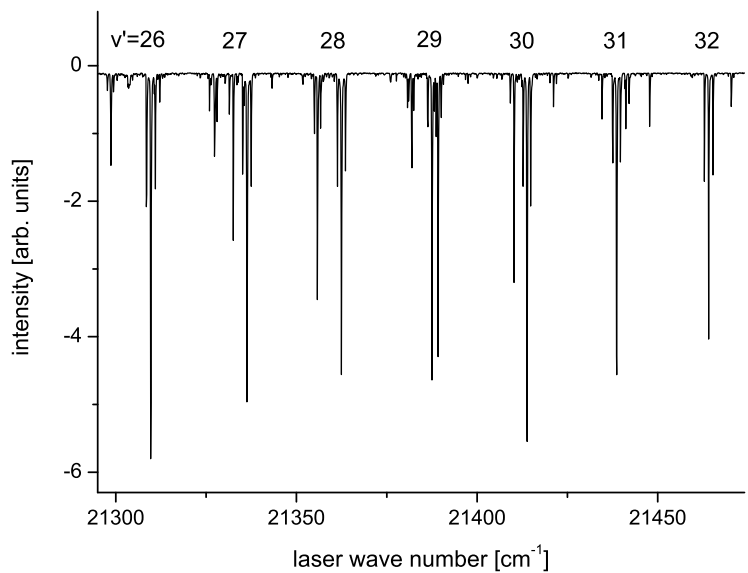}
	\caption{A part of the polarisation spectrum of Cs$_2$ recorded when the ground state level $v''=2, J''=73$ was labelled by the probe laser set at the wave number 12951.237~cm$^{-1}$. The excitation scheme with linearly polarised pump laser light favours Q lines in the spectra, which are then much more pronounced than the P and R lines. Transitions to subsequent $v'$ levels in the 3$^{1}\Pi_{u}$ state are tentatively assigned on top of the drawing, but several extra lines present in the spectrum reveal that the 3$^{1}\Pi_{u}$ state is perturbed by more than one neighbouring state.}
	\label{Fig0}
\end{figure}

\begin{figure}
	\includegraphics[width=1.4\linewidth]{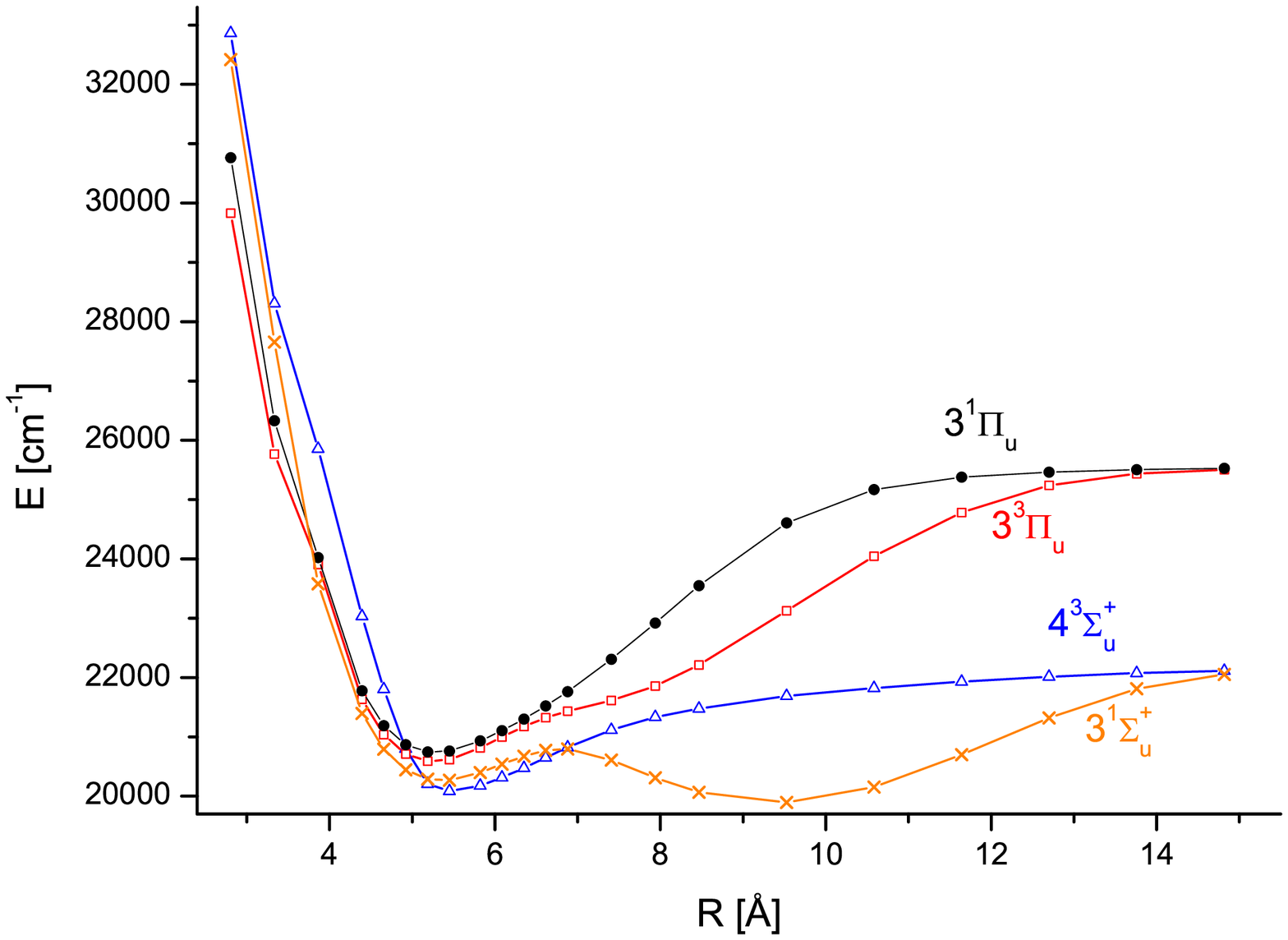}
	\caption{(Colour online) Theoretical potential curves of the four electronic states of Cs$_2$ related to the present experiment~\cite{2}. To guide the eye the calculated points are connected with solid lines.}
	\label{Fig1}
\end{figure}

In the present short communication we report experimental investigation of the 3$^{1}\Pi_{u}$ electronic state of Cs$_2$. Up to now only the lowest vibrational levels $v=0-6$ of this state have been observed~\cite{1} and its sole theoretical description comes from an unpublished thesis by Spies~\cite{2}. Using the V-type double resonance polarisation labelling spectroscopy (PLS) method we were able to extend experimental observations of the 3$^{1}\Pi_{u}$ state up to $v=35$. As our experimental method is described in detail elsewhere~\cite{3,4} we shall not present it here. The only novelty in the present set-up comparing to that described in Ref.~\cite{4} is that as a probe laser we used a home-built single mode external cavity diode laser (ECDL), operating in the range $12000-13200$~cm$^{-1}$ and fixed on selected transitions in the B$^{1}\Pi_{u}$ $\leftarrow$ X$^{1}\Sigma^{+}_{g}$ band system of Cs$_2$~\cite{5,6}. The accuracy of $\pm0.001$~cm$^{-1}$ was achieved by active locking of the laser to High Finesse WS-7 wavemeter. With the pump laser (an excimer laser pumped dye laser on Coumarin 480 dye),  tuneable between 20400 and 21600~cm$^{-1}$, we recorded a few thousand rotationally resolved transitions from the labelled levels in the ground X$^{1}\Sigma^{+}_{g}$ state to levels assigned as $v=0-35$ in the 3$^{1}\Pi_{u}$ state (see an exemplary spectrum in Figure~\ref{Fig0}). Two-step calibration of the spectra using argon and neon spectral lines as well as transmission fringes of a Fabry-P\'{e}rot etalon allowed to determine the wave numbers of the observed spectral lines with an accuracy of about 0.05~cm$^{-1}$. The measured wave numbers were converted to energies of the 3$^{1}\Pi_{u}$ state levels referred to the minimum of the X$^{1}\Sigma^{+}_{g}$ state potential well, using the highly accurate ground state constants~\cite{7}. As the constants reproduce energies of rovibrational levels in the ground state with an accuracy superior to precision of our measurements, no additional errors were introduced into our analysis of the 3$^{1}\Pi_{u}$ state. Our study reveals that instead of regular ladder of levels belonging solely to the 3$^{1}\Pi_{u}$ state we deal with an intricate system of highly perturbed levels belonging to more electronic states. Actually, this can be inferred from a picture of theoretical potential energy curves for the related energy region (Figure~\ref{Fig1}), where the bottom part of the 3$^{1}\Pi_{u}$ state potential runs nearly in parallel with the nearby 3$^{3}\Pi_{u}$ state potential and in addition it is crossed by the 4$^{3}\Sigma^{+}_{u}$ state curve. To simplify the analysis to some extent, we limit our further analysis to $f$ parity levels, prone to perturbations by $\Omega=1$ components of both triplet states. These levels are accessed via Q lines in the spectra, being moreover more pronounced then P and R lines when using our excitation scheme. Our limitation leaves out the E(3)$^{1}\Sigma^{+}_{u}$ state from the analysis as it contains only $e$ parity levels. 

Contrary to the previous report~\cite{1} we find that only the four lowest levels of the 3$^{1}\Pi_{u}$ state, $v=0-3$, are free of observable perturbations and only their positions can be described in a compact way by molecular constants (Table~\ref{table:Tab1}). This suffices to determine the main characteristics of the potential energy curve of the 3$^{1}\Pi_{u}$ state, the equilibrium distance between two caesium nuclei ($R_e=5.27$~\AA), the term energy ($T_e=20684.60$~cm$^{-1}$) and the dissociation energy. It is clear that the molecular 3$^{1}\Pi_{u}$ state dissociates into atoms in 6$^2$S and 7$^2$P states~\cite{1}, however the latter is split into two fine structure components distant by about 181~cm$^{-1}$~\cite{8}. Calculations of Spies~\cite{2} suggest correlation of the 3$^{1}\Pi_{u}$ state with the higher 6$^2$S+7$^2$P$_{3/2}$ asymptote (see Figure~\ref{Fig1}). Using the atomic energy of Cs $E$(7$^2$P$_{3/2}$) = $21946.397 \pm 0.026$~cm$^{-1}$~\cite{8} and dissociation energy of the ground state of Cs$_2$ $\mathscr{D}$(X$^{1}\Sigma^{+}_{g}$) = $3649.5 \pm 0.8$~cm$^{-1}$~\cite{9} we obtain

\begin{equation}
\mathscr{D}(3^1\Pi_u) = \mathscr{D}(\mathrm{X}^1\Sigma^+_g) + E(7^2P_{3/2}) - T_e(3^1\Pi_u) = 4911.3\, \mathrm{cm}^{-1} .
\end{equation}

\begin{table}
  \centering
\caption{ Molecular constants (in cm$^{-1}$) for the 3$^{1}\Pi_{u}$ state of Cs$_2$ representing energies of rovibrational levels in the range $v=0-3$, $J=36-175$, compared with theoretical values~\cite{2}. $\mathscr{D}$ stands for the dissociation energy. In the bottom line the equilibrium distance is also compared. Values in parentheses are uncertainties in units of the last digits, rms stands for the root mean square deviation of the fit.}
\label{table:Tab1} \vspace*{0.5cm}
\begin{tabular}{cccccccc}
\hline
& constant &  experiment & theory &  \\
\hline \
& $T_e$    &  20684.60(3)  & 20734&  \\
& $\omega_e$ &  30.61(1) &     29.3 &      \\
& $B_e$ &  0.009117(30) &     &  \\
& $\alpha_e \times$10$^4$ &  0.445(6)   &          \\
& $D_e \times$10$^8$ &  0.267(9)    &       &      \\
&  rms     &  0.06       &        &        &\\
& $\mathscr{D}$ & 4911.3(9) &   &     &\\
& $R_e$[\AA] &  5.27(1)  &  5.28     &       \\
\hline
\end{tabular}
\end{table} 

\begin{figure}
	\includegraphics[width=1.4\linewidth]{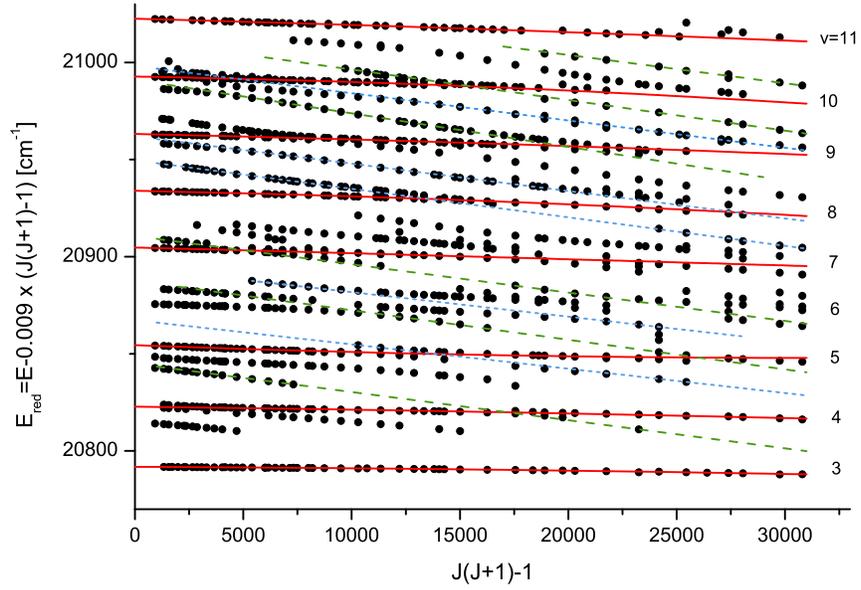}
	\caption{(Colour online) Reduced term values $E_{red}=E-0.09\times [J(J+1)-1]\, [\mathrm{ cm}^{-1}]$ of part of the observed rovibrational levels in the 3$^{1}\Pi_{u}$ $\sim$ 3$^{3}\Pi_{u}$ $\sim$ 4$^{3}\Sigma^{+}_{u}$ complex in Cs$_2$ (dots) plotted against $J(J+1)-1$. The term values calculated from the band constants of the 3$^{1}\Pi_{u}$ state listed in Table~\ref{table:Tab2} are represented by solid lines (red online). The long dashed (green) and short dashed (blue) lines represent approximate term values of the two triplet perturbers.}
	\label{Fig2}
\end{figure}

All vibrational levels of the 3$^{1}\Pi_{u}$ state starting from $v=4$ are heavily perturbed. Energies of part of them are shown in Figure~\ref{Fig2}. The Figure evidently includes both ‘main’ and ‘extra’ levels, the latter belonging to other state(s) inaccessible in direct excitation from the singlet ground state of Cs$_2$ and observable due to mixing of their wave functions with these of the 3$^{1}\Pi_{u}$ state. For several vibrational levels of the 3$^{1}\Pi_{u}$ state we tried to pick out these rotational levels which appeared to be only weakly affected by perturbations and to determine rough band constants by fitting their energies to the formula

\begin{equation}
E(v,J) = T_v + B_v[J(J+1)-1] - D_v[J(J+1)-1]^2 .
\end{equation}

\noindent The results are given in Table~\ref{table:Tab2}. Part of them is displayed by solid lines in Figure~\ref{Fig2}. However, a scatter of the obtained values, e.g. of T$_v$ as a function of $v$, shows that our choice of perturbation-free rotational levels was rather arbitrary. An even simpler formula

\begin{equation}
E(v,J) = T_v + B_v[J(J+1)-1]
\end{equation}

\noindent was applied to levels assumed to represent the perturbers (long dashed and short dashed lines in Figure~\ref{Fig2}). Different slopes of two sets of the corresponding lines confirm our belief that both triplet states 3$^{3}\Pi_{u}$ and 4$^{3}\Sigma^{+}_{u}$ are responsible for the observed interactions. On the other hand the difference is not sufficient to correlate them with particular states.

\begin{table}
  \centering
\caption{Approximate band constants (all values in cm$^{-1}$) of some vibrational levels $v>3$ of the 3$^{1}\Pi_{u}$ state (with no deperturbation attempted).}
\label{table:Tab2} \vspace*{0.5cm}
\begin{tabular}{cccccccc}
\hline
& $v$ &  $T_v$ & $B_v$& $D_v$ $\times10^9$ & rms \\
\hline \
&4	&20822.86	&0.00885	&1.6	&0.09 \\
&5	&20854.72	&0.00850	&-10.1	&0.07 \\
&... & & & & \\
&7	&20904.72	&0.00873	&2.0	&0.11 \\
&8	&20933.89	&0.00878	&6.4	&0.03 \\
&9	&20963.12	&0.00878	&5.0	&0.09 \\
&10	&20992.73	&0.00880	&8.1	&0.06 \\
&11	&21022.51	&0.00871	&2.6	&0.07 \\
&12	&21052.27	&0.00861	&0.6	&0.10 \\
&... & & & & \\
&17	&21193.82	&0.00826	&-1.7	&0.11 \\
&18	&21217.45	&0.00857	&4.5	&0.09 \\
&19	&21246.95	&0.00836	&-0.9	&0.05 \\
&20	&21272.30	&0.00858	&5.7	&0.12 \\
&... & & & & \\
&24	&21378.75	&0.00834	&-0.9	&0.11 \\
&25	&21405.74	&0.00832	&1.3	&0.10 \\
&26	&21432.60	&0.00831	&3.2	&0.12 \\
&27	&21458.81	&0.00834	&4.6	&0.09 \\
&28	&21484.35	&0.00837	&5.1	&0.13 \\
&... & & & & \\
&32	&21587.81	&0.00809	&-1.8	&0.14 \\
&33	&21613.53	&0.00815	&1.8	&0.12 \\
&34	&21639.75	&0.00808	&1.2	&0.09 \\
&35	&21665.53	&0.00806	&0.4	&0.10 \\
\hline
\end{tabular}
\end{table} 

Figure~\ref{Fig2} shows clearly that deperturbation analysis of the strongly interacting 3$^{1}\Pi_{u}$ $\sim$ 3$^{3}\Pi_{u}$ $\sim$ 4$^{3}\Sigma^{+}_{u}$ system requires a coupled channels treatment taking into account all three interacting states and possible spin-orbit and rotational interactions between pairs of them. This is a serious numerical challenge that we have not yet attempted (see an example of such an analysis in Ref.~\cite{10}). This tedious and time consuming work is in future plans of our group, but at present we show the preliminary analysis together with the full list of 3094 experimentally determined term energies of $f$ parity levels belonging to the three electronic states, which may be of interest for the spectroscopic community (see the supplementary data accompanying this paper~\cite{11}).

This work was partially supported by the National Science Centre of Poland (Grant No. 2016/21/B/ST2/02190).

\end{document}